\documentclass[twocolumn,secnumarabic,amssymb, nobibnotes, aps, prb,groupedaddress,superscriptaddress]{revtex4-2}
\usepackage{amsmath,graphicx,latexsym,times,color}
\usepackage{setspace}
\usepackage{hyperref}
\usepackage{array}
\usepackage{textcomp}
\usepackage{titlesec}
\usepackage{physics}
\usepackage{gensymb}
\usepackage{nccmath}
\usepackage{soul} 
\usepackage{empheq} %% loads mathtools, which loads amsmath
\usepackage{comment}

\usepackage{ulem} % STRIKING TEXT
\usepackage[dvipsnames]{xcolor}
\definecolor{blue-violet}{rgb}{0.54, 0.17, 0.89}\newcommand{\V}[1]{\ensuremath{\mathbf{#1}}} %Vector

\let\oldtimes\times  % Make the times "x" use less spacing
\renewcommand\times{{\oldtimes}}

\definecolor{darkpink}{HTML}{4169E1}
\definecolor{darkorchid}{HTML}{bf3eff}

\begin{document}
	\title{Finite-temperature phase transitions across spin symmetries in two-dimensional magnets}

\author{Chen Chen}
\thanks{These authors contributed equally to this work.}
\affiliation{Department of Applied Physics and MIIT Key Laboratory of Semiconductor Microstructure and Quantum Sensing, \href{https://ror.org/00xp9wg62}{Nanjing University of Science and Technology}, Nanjing, China}
\affiliation{\href{https://ror.org/01ahyrz84}{Universit\'e de Toulouse}, \href{https://ror.org/02feahw73}{CNRS}, \href{https://ror.org/03kwnqq69}{CEMES}, Toulouse, France}

\author{Moritz A. Goerzen}
\thanks{These authors contributed equally to this work.}
\affiliation{\href{https://ror.org/01ahyrz84}{Universit\'e de Toulouse}, \href{https://ror.org/02feahw73}{CNRS}, \href{https://ror.org/03kwnqq69}{CEMES}, Toulouse, France}

    \author{Megha Arya}
\affiliation{\href{https://ror.org/01ahyrz84}{Universit\'e de Toulouse}, \href{https://ror.org/02feahw73}{CNRS}, \href{https://ror.org/03kwnqq69}{CEMES}, Toulouse, France}

    \author{Lionel Calmels}
\affiliation{\href{https://ror.org/01ahyrz84}{Universit\'e de Toulouse}, \href{https://ror.org/02feahw73}{CNRS}, \href{https://ror.org/03kwnqq69}{CEMES}, Toulouse, France}

    \author{Yongping Du}
    \email[Contact author: ]{njustdyp@njust.edu.cn}
\affiliation{Department of Applied Physics and MIIT Key Laboratory of Semiconductor Microstructure and Quantum Sensing, \href{https://ror.org/00xp9wg62}{Nanjing University of Science and Technology}, Nanjing, China}
 
	\author{Dongzhe Li}
	\email[Contact author: ]{dongzhe.li@cemes.fr}
\affiliation{\href{https://ror.org/01ahyrz84}{Universit\'e de Toulouse}, \href{https://ror.org/02feahw73}{CNRS}, \href{https://ror.org/03kwnqq69}{CEMES}, Toulouse, France}	
	\date{\today}
	
	\begin{abstract}
Recent advances in intrinsic two-dimensional (2D) magnets have created a need for a unified understanding of finite-temperature phase transitions across different spin-symmetry classes. Here, we investigate finite-temperature magnetism in the classical 2D Heisenberg model using large-scale Monte Carlo simulations, continuously spanning the easy-plane, isotropic, and easy-axis regimes. We find that characteristic temperatures are strongly affected by the interplay of finite-size effects and spin symmetry, induced by magnetocrystalline anisotropy, being most pronounced in the isotropic Heisenberg regime and substantially reduced toward the easy-plane and easy-axis limits. Examining Berezinskii--Kosterlitz--Thouless (BKT) physics in the easy-plane regime, we find that the characteristic temperatures independently estimated from the magnetization and heat capacity deviate significantly from those proposed by BKT theory, but match a previously reported anomaly in the XY-model. We report that characteristic temperatures instead are correlated with the proliferation of vortex-antivortex pairs, suggesting the gain in entropy by their accelerated nucleation as the origin of the anomaly. Establishing this connection both theoretically and numerically, we argue that the thermodynamic characteristic temperature in finite easy-plane Heisenberg magnets, including the deviation from BKT theory, is subject to the interplay between finite-size spin correlations and vortex proliferation. These results clarify how experimental and technologically relevant temperature scales evolve with magnetic anisotropy, spin symmetry, and system size in finite 2D magnets.
	\end{abstract}
	
	\maketitle

\section{INTRODUCTION}

The magnetic order at finite temperatures in two-dimensional (2D)
magnets is jointly determined by spin symmetry and magnetic anisotropy. According to the Hohenberg-Mermin–Wagner theorem (HMW), a 2D isotropic Heisenberg model with short-range exchange interactions cannot sustain true long-range magnetic order at finite temperatures~\cite{mermin1966absence,hohenberg1967existence}. 
To obtain magnetism in the 2D limit, magnetic moments and ordering have been induced or engineered through extrinsic mechanisms, such as defects, vacancies, adatoms, or doping~\cite{yazyev2010emergence,tongay2012magnetic}. These approaches provide routes to magnetism in otherwise nonmagnetic 2D systems. A major breakthrough came in 2016--2017 with the first experimental demonstrations of intrinsic magnetism in atomically thin van der Waals (vdW) materials~\cite{lee2016ising,gong2017discovery,huang2017layer}, opening new opportunities for exploring magnetic phenomena in reduced dimensionality \cite{gibertini2019magnetic,RMP_2dmagnetism}. In these materials, magnetic anisotropy can break continuous spin rotational symmetry, effectively suppressing thermal fluctuations and thereby stabilizing magnetic order at finite temperatures. Recent experiments further demonstrate that changes in magnetic anisotropy can drive a crossover from 3D Heisenberg to 2D Ising magnetism in a vdW magnet of Fe$_3$GeTe$_2$~\cite{xiao2026anomalous}.
Additionally, 2D magnets with strong easy-plane~\cite{cheon2025nature} and six-clock anisotropy~\cite{gao2026six}, as well as short-range quasi-2D XY systems in which critical fluctuations qualitatively alter the ordering behavior~\cite{hu2026}, highlight the different mechanisms governing finite-temperature phase transitions in 2D magnets.

\begin{figure*}[t]%---------------------------------------------------------------------------------------------------------------------FIGURE 1
	\centering
	\includegraphics[width=1.0\linewidth]{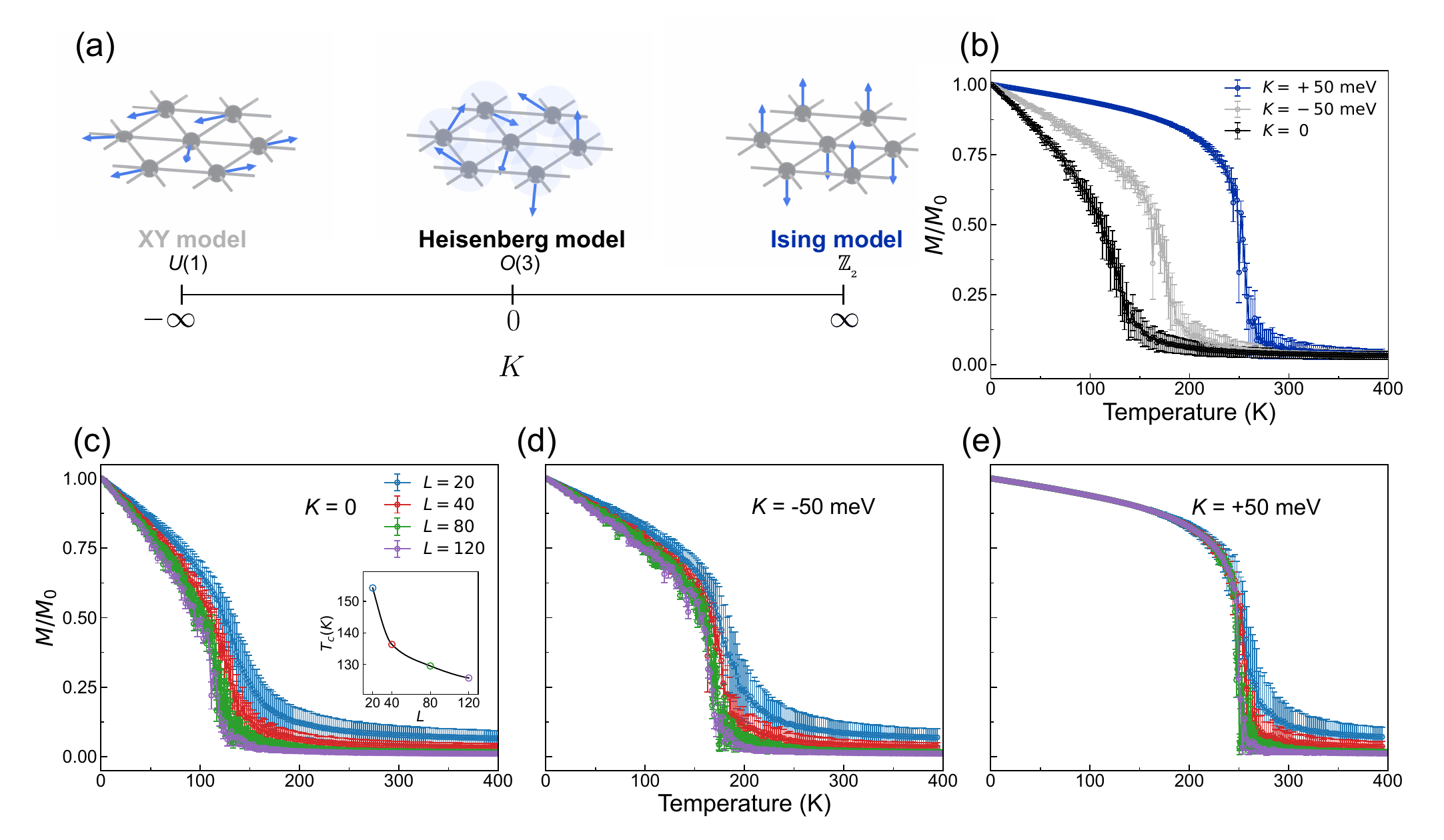}\\
	\caption{\label{Fig1} (a) Symmetry limits of the 2D anisotropic Heisenberg model as a function of the magnetocrystalline anisotropy $K$. The limits $K\rightarrow-\infty$, $K=0$, and $K\rightarrow+\infty$ correspond to the XY model with $U(1)$ symmetry, with spins confined to the plane, the isotropic Heisenberg model with $O(3)$ symmetry, with unrestricted spin orientations, and the Ising model with $\mathbb{Z}_2$ symmetry, with spins restricted to two opposite orientations, respectively. (b) Temperature dependence of the normalized magnetization $M/M_0$, with $M_0=M(0)$, for representative anisotropy values on a lattice with $40\times40$ sites, for three representative values of $K$. The deviation between the curves illustrates the anisotropy-mediated deviations from the finite-size-induced magnetization at $K=0$. (c-e) Temperature dependence of the normalized magnetization, $M/M_0$, for $K=0$, $K=-50$ meV, and $K=+50$ meV. Results are shown for different system sizes to illustrate differences in the response to finite-size effects depending on the respective symmetry group of the system. The inset in (c) indicates the progression of the respective critical temperature $\lim_{L\to\infty}T_c(L)=0~$K for $K=0~$meV, in accordance with the HMW theorem. 
    }
    
\end{figure*}%------------------------------------------------------------------------------------------------------------------------------	

%%%% FIGURE 2
\begin{figure*}[t]
	\centering
	\includegraphics[width=1.0\linewidth]{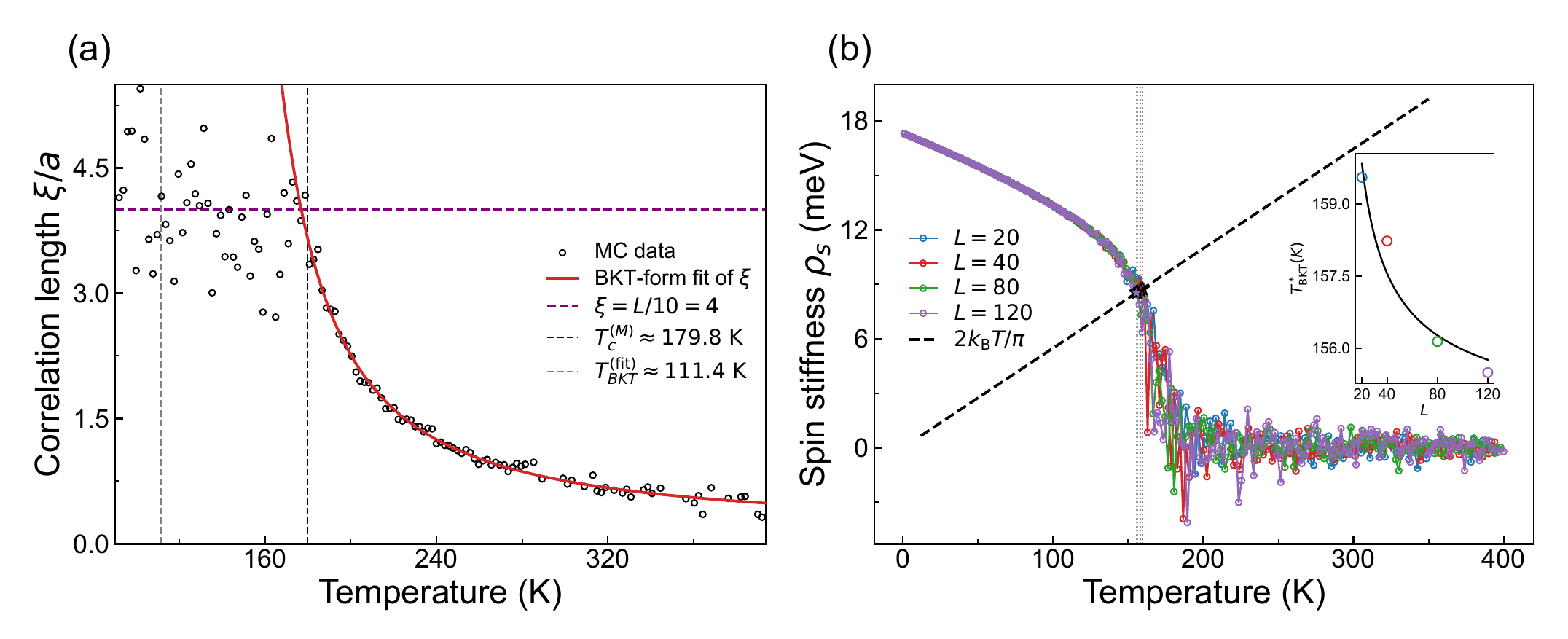}\\
	\caption{\label{Fig3} {Correlation-length and spin stiffness signatures in the easy-plane regime.}
    (a) Temperature dependence of the in-plane correlation length $\xi/a$ for $K=-50~\mathrm{meV}$. The red solid line shows the fit by Eq.~(\ref{eq:bkt_correlation_length}) in the high-temperature regime, which diverges at $T_{\text{BKT}}$ (light grey dashed line). The black vertical dashed line indicates $T_c^{(M)} \simeq 179.8~\mathrm{K}$ obtained from the magnetization $M(T)$. The purple horizontal dashed line denotes $\xi=L/10$, indicating that the magnetization transition is associated with a correlation length comparable to the system size. 
    (b) Area-normalized spin stiffness $\rho_s$ for different system sizes $L$. The black dashed line denotes the universal-jump criterion $2k_{\mathrm{B}}T/\pi$, and the colored vertical dotted lines mark the corresponding finite-size crossings. The inset shows the fit to Eq.~(\ref{bkt_fit}) with $L_0=1$, yielding an estimated $T_{\text{BKT}}^*\simeq153.1$ K.
    }
\end{figure*}

Another important aspect is the role of finite-size effects. Although the HMW theorem excludes long-range magnetic order at finite temperature in the thermodynamic limit~\cite{mermin1966absence,hohenberg1967existence}, this limit is far removed from experimentally relevant length scales. In practical applications, real 2D magnets and spintronic devices are necessarily finite, typically extending only to the micrometer scale \cite{kim2019micromagnetometry,wahab2021quantum}. Recent large-scale simulations have shown that, even in the absence of magnetic anisotropy, isotropic exchange interactions can sustain short-range magnetic order at finite temperatures in finite 2D systems~\cite{jenkins2022breaking}. When the spin correlation length becomes comparable to or larger than the system size, the spins remain correlated across the sample, giving rise to a finite crossover temperature. Since this temperature decreases only slowly with increasing system size, finite-size magnetic ordering can persist up to experimentally relevant micrometer length scales.

Here, we use the anisotropic 2D Heisenberg model to investigate how finite-temperature behavior changes across the easy-plane, isotropic, and easy-axis regimes as the magnetic anisotropy continuously modifies the spin symmetry. We first examine how finite-size effects depend on symmetry, showing that they are strongest in the isotropic Heisenberg regime and progressively reduced toward the easy-plane and easy-axis limits. In particular, we revisit the well-established Berezinskii-Kosterlitz-Thouless (BKT) transition in the easy-plane regime~\cite{berezinskii1971destruction,kosterlitz1973ordering}, examining how BKT physics and finite-size magnetic correlations give rise to different characteristic temperatures. By combining magnetization, heat-capacity, correlation-length, and spin-stiffness analyses, we show that the thermodynamic anomaly occurs at a temperature well above $T_{\text{BKT}}$ obtained from the universal-jump criterion, demonstrating that the two temperature scales cannot be directly identified. Finally, through a direct analysis of vortex excitations and a vortex-occupancy entropy, we show that the thermodynamic anomaly is closely associated with the rapid proliferation of vortices, pointing to an interplay between finite-size correlations and vortex excitations rather than $T_{\text{BKT}}$.

%===============================================
% SECTION: MODEL AND METHODS
%===============================================
\section{MODELS AND NUMERICAL METHODS}
In this work, all MC simulations are based on the Metropolis algorithm, as implemented in the \textsc{spinaker} code and used in Refs.~\cite{li2024prediction, kollwitz2026entropy} [see Appendix A for computational details], and the following Hamiltonian
\begin{equation}\label{eq:hamiltonian}
H(\mathbf{m};J,K)  =-J\sum_{\langle ij\rangle}(\V{m}_i \cdot \V{m}_j)-K\sum_{i=1}^N (m_i^z)^2~.
\end{equation}
Here, the spin configuration $\mathbf{m}=(\mathbf{m}_1,...,\mathbf{m}_N)$ is represented by classical unit vectors $\V{m}_i$ on a 2D triangular lattice with sites $\V{r}_i$, which is a representative lattice for 2D vdW magnets and provides sixfold symmetry and a high coordination number. Here, we used $J=5$ meV/atom as the nearest-neighbor (NN) exchange coupling, which favors a ferromagnetic alignment, and $K$ is the uniaxial magnetocrystalline anisotropy energy (MAE). For simplicity, we focus on a minimal model that retains $J$ and $K$. The Dzyaloshinskii-Moriya interaction is neglected, as we have verified that it has only a minor effect on the phase transitions considered in this work, while other contributions, including dipolar interactions, beyond NN exchange, and higher-order exchange interactions, are assumed to be subdominant.
As illustrated in Fig.~\ref{Fig1}(a), the Hamiltonian interpolates between three limiting cases depending on the value of $K$, each of which belongs to a different symmetry class: 
\begin{itemize}
    \item[(i)] For $K=0$ the Heisenberg model is isotropic with $O(3)$-rotational symmetry. This continuous symmetry prevents the occurrence of long-range magnetic order in infinite systems, according to HMW. However, finite systems exhibit a pronounced magnetic crossover when the spin-correlation length becomes comparable to the system size~\cite{prb_2025_heisenberg,jenkins2022breaking}.
    \item[(ii)] The case $K<0$ produces easy-plane anisotropy that favors a spin orientation within the surface plane with $m_z=0$. The resulting $U(1)$-symmetry still hinders the formation of long-range order according to the HMW theorem, but inherits the possibility of finite-size induced ordering from the isotropic case. However, in the limit $K\rightarrow-\infty$, the system approaches the XY model, which exhibits quasi-long-range magnetic order due to the BKT mechanism, a phenomenon that has also been reported for the Heisenberg model~\cite{costa2003phase,prl_2021_bkt,prb_2021_bkt,liao2024tunable}.
    \item[(iii)] In the case of an easy-axis with $K>0$ all continuous symmetries are broken, leaving only the spin-flip symmetry $\mathbb{Z}_2$ and thus allowing for long-range order. In the strong easy-axis limit $K\rightarrow\infty$, the system develops Ising-like behavior~\cite{PRB_1976,torelli2019calculating}. 
\end{itemize}
For comparability to realistic systems and reasons of numerical stability, we investigate the cases above in the limits $K\in[-50, 50]~$meV, which thus are bounded by $|K|\leq10J$. For larger $K$, the Metropolis algorithm returns poorly sampled observables, due to high energy penalty for small deviations from spin directions that are preferred by MAE.

%===============================================
% SECTION: EMERGENCE OF MAGNETIC ORDER
%===============================================
\section{RESULTS}
\subsection*{A. Emergence of magnetic order}
The effect of these symmetry classes on the thermal position of magnetic phase transitions can be observed in Fig.~\ref{Fig1}(b), where the temperature-dependent magnetization 
\begin{equation}\label{eq:magnetization}
    M(T) = \frac{1}{N}\left\langle\left|\sum_{i=1}^N \mathbf{m}_i \right|\right\rangle~,
\end{equation}
averaged from MC simulation on a lattice with $40\times40$ sites, for $K=0,\pm50$~meV is presented. While the thermal position of the phase transition for $K=0$ is purely determined by the temperature at which the correlation length is on the scale of the system size (cf. Ref.~\cite{jenkins2022breaking}), the cases $K=\pm50~$meV deviate significantly, suggesting fundamentally different microscopic origins of these transitions. This presumption is supported by the system-size-dependent magnetization curves shown in Fig.~\ref{Fig1}(c-e). Especially the case of $K = 0$ is extremely sensitive to size, eventually reaching a critical temperature of $T_c=0~$K [cf. inset in Fig.~\ref{Fig1}(c)] at the size of the observable universe \cite{jenkins2022breaking} or, more conservatively, at the size of Texas \cite{Bramwell1994magnetization}. For $K=\pm50~$meV, the finite-size effect is less pronounced. Aside from different magnitudes in the dependence of the magnetization to the system size, it has to be noted that the deviations among the $M(T)$ curves for $K=0$, $K=-50$, and $K=+50$ meV observed in Fig.~\ref{Fig1}(b) persist over a wide range of lattice dimensions. Thus, the origin of the deviations lies beyond finite-size effects.
To distinguish these finite-size effects from signatures of BKT physics, in the following subsections we review the relevant BKT criteria and examine the corresponding thermodynamic observables in our simulations.

%%%% FIGURE 3
\begin{figure*}[t]
	\centering
	\includegraphics[width=0.9\linewidth]{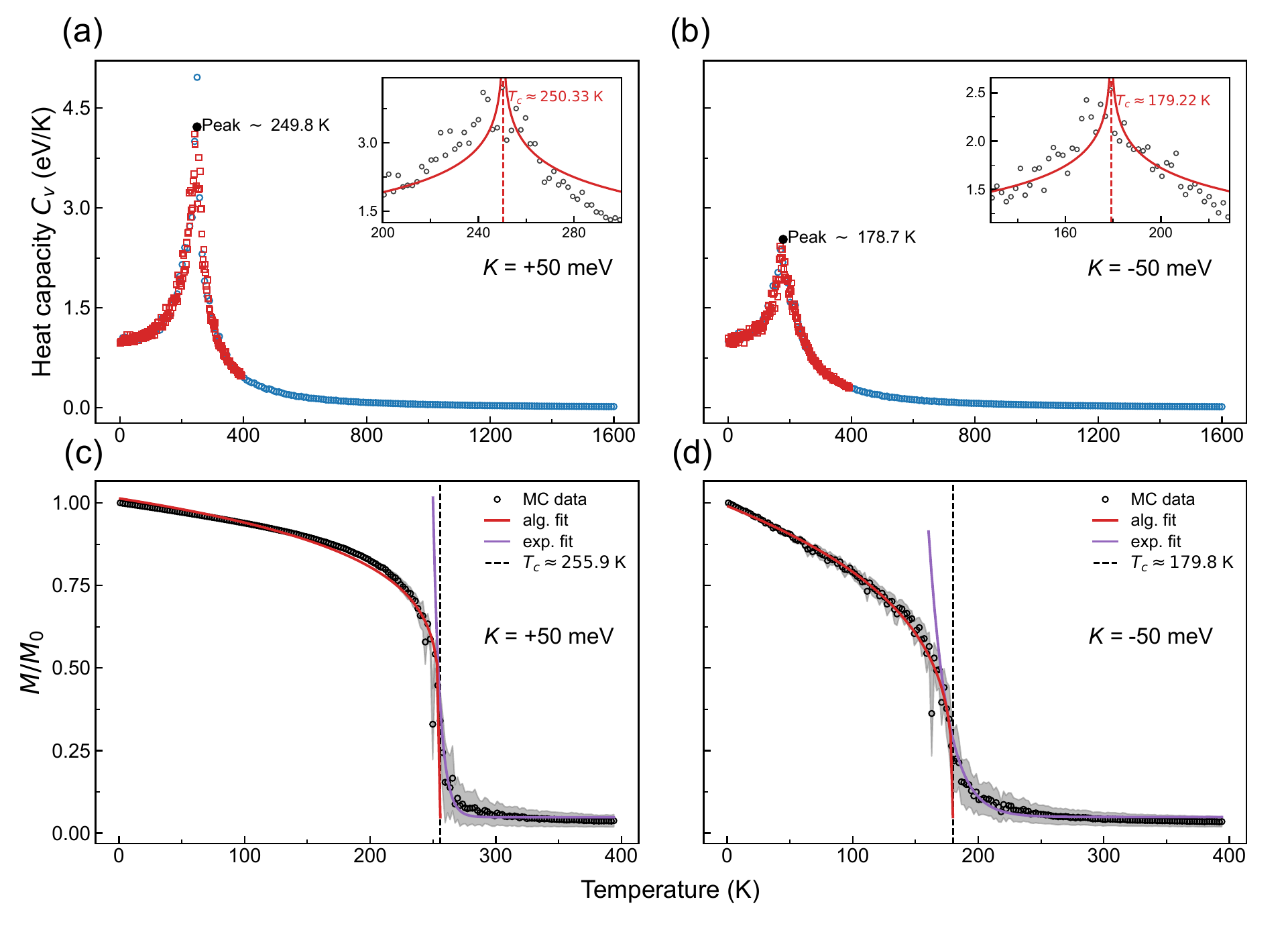}
	\caption{\label{Fig2} Determination of the critical temperature from heat-capacity and magnetization analyses. (a-b) Heat capacity $C_V$ as a function of temperature for $K=+50$ meV and $K=-50$ meV, respectively. Insets show fits by the power law in Eq.~(\ref{eq:power_law_heat_capacity}) to the peak region used to extract the critical temperature. (c-d) Corresponding temperature dependence of the normalized magnetization $M(T)/M_0$.
    The low-temperature regime is fitted using an algebraic function, whereas the high-temperature regime is described by an exponential function. The resulting fits (red lines) provide an estimate of the critical temperature obtained from the magnetization. The critical temperatures extracted from the two independent approaches are in good agreement.  
    }
\end{figure*}

%===============================================
% SECTION: BKT THEORY
%===============================================
\subsection*{B. Topological phase transitions}
Aside from long-range magnetic order, which occurs in systems with spontaneously broken symmetry in accordance with the HMW theorem, systems with unbroken symmetries can feature quasi-long-range order due to the BKT mechanism. This reflects that long-range excitations, which would destroy any magnetic order in sufficiently large systems, are screened by thermally nucleated vortex-antivortex pairs \cite{berezinskii1972destruction_QE,kosterlitz1973ordering,Kosterlitz1974}. 
This screening leads to a divergence of the spin-correlation length $\xi$ as the critical temperature $T_{\text{BKT}}$ is approached from above, which takes the
form~\cite{berezinskii1971destruction,kosterlitz1973ordering,Kosterlitz1974}
\begin{equation}\label{eq:bkt_correlation_length}
T_{\text{BKT}}<T:\quad
\xi(T)=\xi_0
\exp\left[
\frac{b}{\sqrt{T/T_{\text{BKT}}-1}}
\right],
\end{equation}
where $\xi_0$ and $b$ are nonuniversal constants and
$T_{\text{BKT}}$ is the BKT transition temperature. Below
$T_{\text{BKT}}$, the system exhibits quasi-long-range order
characterized by algebraically decaying spin correlations.

A fit of Eq.~(\ref{eq:bkt_correlation_length}) to the numerically obtained correlation length in Fig.~2(a) reveals good agreement between the BKT form and the simulation results. 
Details of the calculation and fitting of the correlation function are provided in Appendix~B. This agreement implies that the system described by the Hamiltonian in Eq.~(\ref{eq:hamiltonian}) for $K=-50~$meV could belong to the BKT universality class \cite{liao2024tunable}. However, this assignment is not conclusive because exponential correlation decay can also occur in the pseudocritical regime of finite systems \cite{tomita2014finite}. The connection to finite-size effects becomes apparent from the horizontal line in Fig.~\ref{Fig3}(a), which indicates that a finite magnetization occurs for a correlation length of $\xi \approx L/10$, which is thus on the same order of magnitude as the system dimension. A BKT origin of the magnetization is further refuted by the obtained critical temperature of $T_{\text{BKT}}= 111.4~$K, which is significantly lower than the critical temperature $T_c=179.8~$K obtained from magnetization. Nevertheless, it establishes the connection between the suppression of the magnetization in the easy-plane magnet [cf. Fig.~\ref{Fig1}(e)] and the loss of quasi-long-range correlations [cf. Fig.~\ref{Fig3}(a)].

A more reliable method to determine $T_{\text{BKT}}$, which not only builds on high-temperature behavior, is the Nelson--Kosterlitz universal-jump condition~\cite{nelson1977universal}, which states that the renormalized spin stiffness $\rho_s(T,L)$ near $T_{\text{BKT}}$ for any system in the BKT universality class has to satisfy
\begin{equation}\label{eq:universal_jump_condition}
    \lim_{T\to T_{\text{BKT}}^*(L)} \frac{\rho_s(T,L)}{k_{\text{B}}T}=\frac{2}{\pi} ~.
\end{equation}
This criterion returns the temperature $T_{\text{BKT}}^*(L)$ at which it is entropically favored for the vortex-antivortex pairs to dissolve into isolated vortices \cite{berezinskii1972destruction_QE,kosterlitz1973ordering,Kosterlitz1974} (see Sec.~E, ``Vortex Analysis", for details). The central quantity is the spin stiffness, which describes the energetic resistance of the system to a long-wavelength twist, or equivalently, the curvature of the free energy $F$ with respect to a distortion $\mathbf{q}$. At the $\overline{\Gamma}$-point, $\mathbf{q}=\boldsymbol{0}$, it takes the computable form \cite{obuchi2012spin}
\begin{equation}\label{eq:spin_stiffness}
    \rho_s = \frac{1}{NA}\left. \frac{\partial^2F(\mathbf q)}{\partial q_{\alpha}^2} \right|_{\mathbf q=\boldsymbol{0}} = \frac{\langle B_{\alpha}\rangle - \beta\text{Var}(I_{\alpha})}{NA}~,
\end{equation}
with $\beta^{-1}=k_{\text{B}}T$, $\alpha\in\{x,y\}$ and the area of the hexagonal unit cell $A=\sqrt{3}a^2/2$ with normalized lattice constant $a=1$ (see Appendix~C for details). 
The numerically estimated spin stiffness for $K=-50~$meV and various system sizes $L\times L$ is shown in Fig.~\ref{Fig3}(b). Here, it should be noted that the offset at $T=0~$K coincides with the typical micromagnetic spin stiffness, which is typically obtained from atomistic exchange in a long-wavelength approximation~\cite{rybakov2022magnetic,zhu2026beating}, and for a hexagonal NN model, like in our case, it takes the value 
\begin{equation}
    \rho_s(0) = \frac{1}{2A}\sum_{i}J_{ij} r_{ij}^2\big|_{j=0} = 2\sqrt{3}J~.
\end{equation}
Here, $r_{ij}^{2}=(r_{ij}^{x})^{2}+(r_{ij}^{y})^{2}$ denotes the squared
bond length, while $r_{ij}^{\alpha}$ in Eq.~(\ref{eq:stiffness_terms}) represents the component
of the bond vector along the twist direction $\alpha$.

The estimation of size-dependent $T_{\text{BKT}}^*(L)$ from the universal-jump condition in Eq.~(\ref{eq:universal_jump_condition}) comes down to determining the intersection between the spin stiffness and a straight line of the form $f(T)=2k_{\text{B}}T/\pi$, which is displayed as a dashed line in Fig.~\ref{Fig3}(b). Further, taking into account the renormalization of the stiffness with the system size, which, by theory, follows \cite{Bramwell1994magnetization}
\begin{equation}\label{bkt_fit}
    T_{\text{BKT}}^*(L) = T_{\text{BKT}} +\frac{C}{\left[\ln\left(L/L_0\right)\right]^2}~,
\end{equation}
with constants $C, L_0$, leading to an estimate of $T_{\text{BKT}}\approx158~$K. This characteristic scale is moderately lower than the apparent critical temperature of $T_c = 179.8~\mathrm{K}$ inferred from the magnetization and heat capacity, which will be discussed in the upcoming subsection. Although the difference is consistent with the pronounced size-dependence of the magnetization in a 2D easy-plane system, which is determined by the correlation between system size and spin correlation length (cf. Eq.~(\ref{eq:bkt_correlation_length})), the deviation is important for accurately identifying critical temperatures in real materials.

\begin{figure*}[t]%---------------------------------------------------------------------------------------------------------------------FIGURE 4
	\centering
    \includegraphics[width=1.0\linewidth]{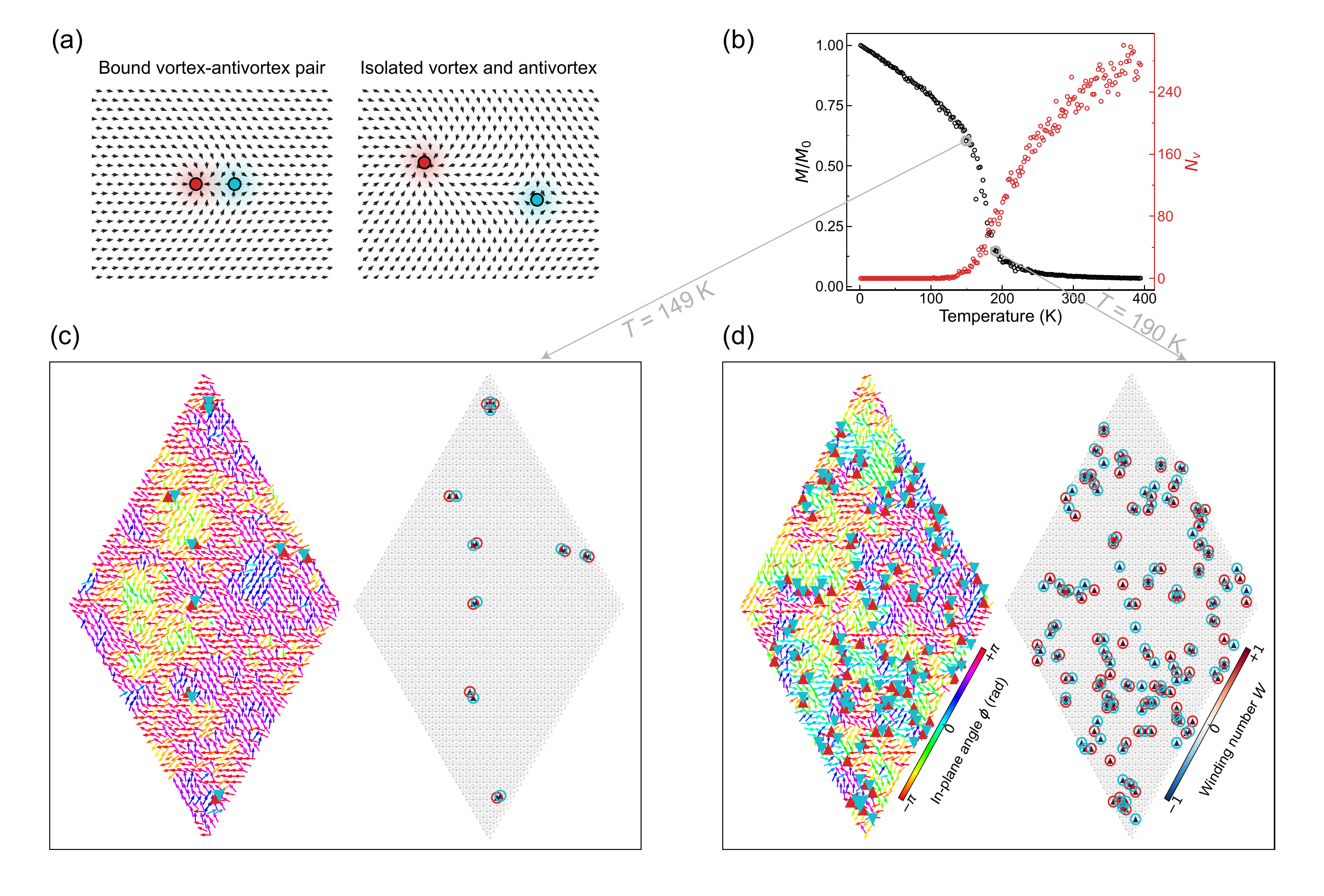}\\
	\caption{\label{Fig4} Vortex excitations in the easy-plane regime for $K=-50$ meV in a $40\times 40\times 1$ triangular lattice. (a) Schematics of spin configurations of a bound vortex-antivortex pair (left) and an isolated vortex and antivortex (right). The color bar indicates the winding number $\nu=1$ (red) and $\nu=-1$ (cyan). (b) Temperature dependence of the normalized magnetization $M/M_0$ (black, left $Y$-axis) and vortex number $N_v$ (red, right $Y$-axis). Two representative temperatures, $T=149$~K (c) and $T=190$~K (d), are selected. At each temperature, the spin texture is projected onto the $xy$ plane and colored according to the azimuthal angle $\phi$, together with the corresponding vortex distribution.
    }
\end{figure*}%-----------------------------------------------------------------------------------------------------------------------------

%===============================================
% SECTION: HEAT CAPACITY AND MAGNETIZATION
%===============================================
\subsection*{C. Characteristic temperatures from thermodynamic anomalies}

For practical applications, it is important to establish whether different thermodynamic observables identify a common characteristic temperature in finite magnetic systems. It is therefore important to understand whether conventional thermodynamic criteria yield the same temperature scale for real materials as the BKT analysis discussed in the previous subsection.
Next to the magnetization [cf. Eq.~(\ref{eq:magnetization})], another important thermodynamic observable is the heat capacity, which we compute from thermal fluctuations of the internal energy $E$ as
\begin{equation}\label{eq:heat_capacity}
    C_V = \frac{\text{Var}(E)}{N k_{\text{B}} T^2}~.
\end{equation}

Fig.~\ref{Fig2}(a-b) show the temperature-dependent heat capacity for the extreme cases of $K=\pm50~$meV. The insets show zooms on the position of the presumptive phase transition. The red line is the result of a least-squares fit using the usual power-law function near a critical temperature
\begin{equation}\label{eq:power_law_heat_capacity}
    C_{\nu}(T)=A\,|T-T_c|^{\eta}+C~.
\end{equation}
As already observed in Fig.~\ref{Fig1}(d-e) both setups return well distinguished transition temperatures $T_c(K)$ temperatures of $T_c(50~\text{meV})\approx250~$K and $T_c(-50~\text{meV})\approx179~$K. These transition temperatures are in good agreement with those obtained from the magnetization, as shown in Fig.~\ref{Fig2}(c-d). Here we employed a more sophisticated fitting procedure, which not only considers the algebraic power-law behavior known for Ising-like and BKT phases at low temperatures, but also the exponential decay at high temperatures, as expected for pseudocritical and BKT transitions \cite{tomita2014finite}
\begin{equation}\label{eq:fit_magnetization}
    M(T)= \begin{cases} A(T_c-T)^{\eta}+C, & T \le T_c-\delta T \\[2pt] A\,f(T)+C, & T > T_c - \delta T
    \end{cases}
\end{equation}
with critical exponent $\eta$ and shift $\delta T$ between $T_c$ and the temperature at which algebraic and exponential decay coincide. To assure continuous differentiability of $M$ at $T_c-\delta T$ the function $f(T)$ is chosen as \cite{kollwitz2026entropy}
\begin{equation}
    f(T)= (\delta T)^{\eta} \exp\!\left[ \frac{\eta}{\delta T}(T_c-\delta T - T) \right]~.
\end{equation}
Since by design this function is applicable to any magnetization curve in a finite Heisenberg magnet, regardless of the origin being Ising-like, pseudocritical, or BKT, it highlights the difficulty of distinguishing these mechanisms based on thermodynamic observables.
For the easy-axis case of $K=+50$ meV, the resulting characteristic temperature from magnetization, $T_\text{c}\simeq255.9$ K, is consistent with that extracted from the heat capacity. The easy-plane case, however, reveals an important deviation. For
$K=-50~\text{meV}$, the heat capacity and magnetization identify
essentially the same characteristic temperature,
$T_{C_V}\simeq T_M\simeq179~\text{K}$, whereas the
universal-jump analysis of the spin stiffness gives
$T_{\text{BKT}}\simeq153~\text{K}$.
Interestingly, a heat-capacity anomaly above $T_{\text{BKT}}$ has also been reported for the XY model~\cite{nguyen2021superfluid}.
On top of that, it has to be noted that the sharp peak in the heat capacity that we observe in our simulations [cf. Fig.~\ref{Fig2}(b) and (d)] is in disagreement with the broader peaks associated with the BKT transition \cite{nguyen2021superfluid, mitrovic2010monte, nguyen2020investigation, tutsch2014evidence}. Taken together, this raises the question of whether the critical temperature obtained for easy-plane Heisenberg magnets can be associated with a BKT transition at all. Moreover, the similar thermodynamic signatures of phase transitions associated with different underlying mechanisms \cite{tomita2014finite} -- finite-size induced, ordered, or BKT -- further indicate the ambiguity of fingerprints for BKT transitions. In this spirit, we develop and apply a vortex-counting statistic to investigate the connection between phase transitions and topological excitations, as well as the origin of the thermodynamic anomaly.

\begin{figure*}[tp]
	\centering
	\includegraphics[width=1.0\linewidth]{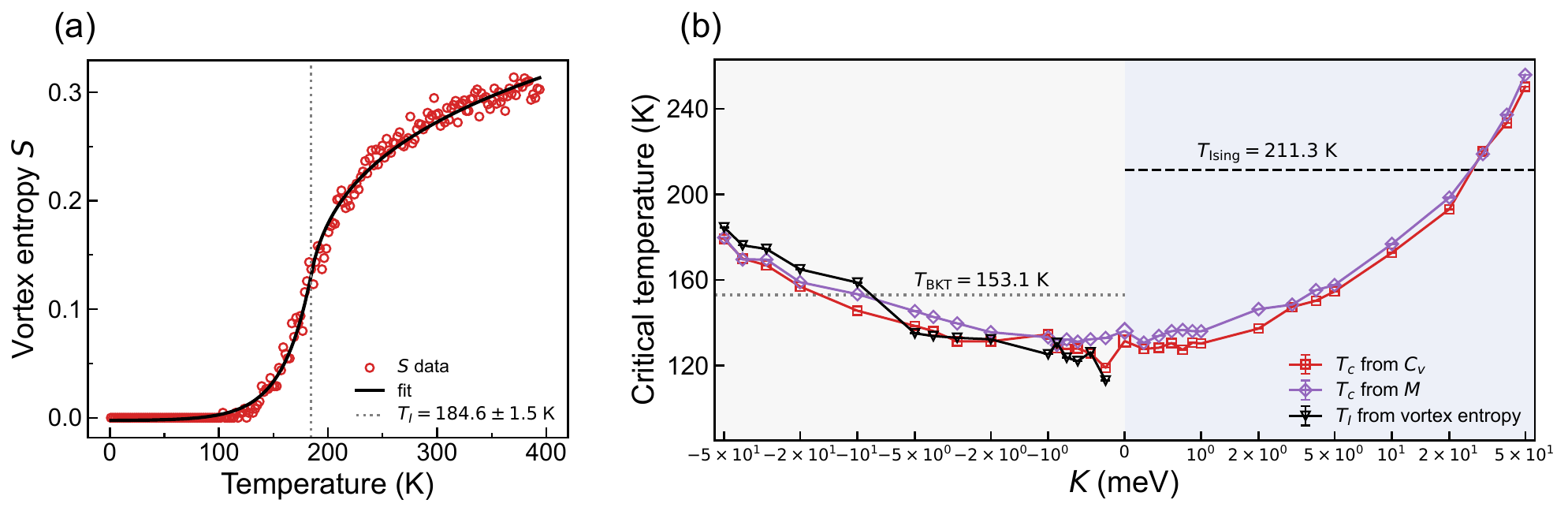}
	\caption{\label{Fig5} (a) Temperature dependence of the vortex entropy $S$ for $K=-50~$meV obtained from a $40\times 40$ triangular lattice. The solid line represents the fit to the mirrored Eq.~(\ref{eq:fit_magnetization}). The vertical line marks the inflection point $T_I$. (b) Critical temperatures as a function of $K$ obtained from heat capacity $C_V$ (red), magnetization $M$ (violet), and vortex entropy $S$ (black), computed on $40\times 40$ lattices. The critical temperatures of the BKT [cf. Fig.~\ref{Fig3}(b)] and Ising models are shown as horizontal dashed lines. 
    }
\end{figure*}
    
\subsection*{D. Vortex-antivortex pairs in BKT theory} 
To illustrate the topological origin of the BKT transition, we assume the continuous magnetization $\mathbf{m}~:\mathbb{R}^2\to\mathbb{S}^n$ that maps a position $\mathbf{r}\in\mathbb{R}^2$ on a 2D lattice to a position on the sphere $\mathbb{S}^n$ and, with respect to the polarized state, has the exchange energy
\begin{equation}\label{eq:continuum_exchange}
    E(\mathbf{m}) = \rho_s(T)\int_{\mathbb{R}^2} |\nabla\mathbf{m}|^2~\mathrm{d}\mathbf{r} ~.
\end{equation}
For the case of the Heisenberg model with $n=2$, the second homotopy group is $\pi_2(\mathbb{S}^2)=\mathbb{Z}$, which states that topologically distinct configurations coexist. However, the energy of a configuration only has a lower bound, as shown in Ref. \cite{belavin1975metastable}
\begin{equation}
    E(\mathbf{m}) \geq 4\pi\rho_s |Q|~,\quad Q=\frac{1}{4\pi}\int_{\mathbb{R}^2} \mathbf{m}\cdot\left(\frac{\partial\mathbf{m}}{\partial r_1}\times\frac{\partial\mathbf{m}}{\partial r_2}\right)~\mathrm{d}\mathbf{r}~,
\end{equation}
with $Q\in\mathbb{Z}$ being the topological invariant of the Heisenberg model. Thus, regarding Eq.~(\ref{eq:continuum_exchange}), there exists no energy barrier between adjacent topological sectors and therefore no topological phase transition \cite{kosterlitz2018topological}. Only when additional interactions, such as Dzyaloshinskii-Moriya or frustrated exchange interactions, are considered can topologically distinct states, such as skyrmions~\cite{bogdanov1994thermodynamically}, be stabilized. For the XY model with $n=1$, on the other hand, this energy barrier is, in principle, infinite, consequently allowing for topologically distinct phases followed by a phase transition.

Topologically distinct objects in the XY model are vortices, which are characterized by a whirling magnetization $\mathbf{m}(\varrho, \phi)=(\cos\Phi(\phi), \sin\Phi(\phi), 0)$ that, due to $\Phi(\phi)=\nu\phi$, rotates $\nu$ times around its center. In small-angle approximation $\Phi\to0$ this profile inserted in Eq.~(\ref{eq:continuum_exchange}) gives an isolated vortex energy of 
\begin{equation}\label{eq:vortex_energy}
E_v = \pi\nu^2\rho_s(T) \ln\left(\frac{L}{\varepsilon}\right),
\end{equation}
integrated over a $L\times L$-system and with $\varepsilon$ being the diameter of the singular vortex core area around $\varrho=0$.
In such a system, there exist $Z=(L/\varepsilon)^2$ possible vortex positions, giving rise to a configurational entropy $S_{v}=k_{\text{B}} \ln Z$ for a single vortex. 
Taken together, the free energy of a micro-canonical ensemble of systems that contain exactly one (anti-)vortex with $\nu=\pm1$ becomes \cite{kosterlitz1973ordering,nelson1977universal}
\begin{equation}
    F_{v}=E_{v}-TS_{v} = \left[\pi\rho_s(T) - 2k_{\text{B}}T\right]\ln\left(\frac{L}{\varepsilon}\right)~.
\end{equation}
Compared to the universal jump condition (Eq.~(\ref{eq:universal_jump_condition})), this states that for $T>T^*_{\text{BKT}}(L)$ the entropic term dominates the reduction of the free energy. 
 In this picture $T^*_{\text{BKT}}(L)$ is the critical temperature for a transition between a phase of ordered vortices and an entropy-driven vortex-plasma phase and thus associated with the unbinding of vortex-antivortex pairs, as illustrated in Fig.~\ref{Fig4}(a).

To examine this microscopic picture behind the BKT transition, we compute the average vortex-pair number $N_{v}$ numerically from
Monte Carlo simulations.
By iterating over all triples $\langle ijk\rangle$ of neighboring lattice sites, we identify a vortex numerically by its discretely sampled vortex number 
\begin{equation}
    \nu_{ijk}= \frac{1}{2\pi} \left(\Phi_{ij} +\Phi_{jk} +\Phi_{ki}\right) = \left\{\begin{array}{cc}
        1 & \text{vortex} \\
        -1 & \text{antivortex}\\
        0 & \text{trivial}
    \end{array}\right.
\end{equation}
with $\Phi_{ij}=\Phi_j-\Phi_i$, which, in reference to the XY-model, is determined purely from the easy-plane component of spins as $\Phi_i=\operatorname{atan2}(m_i^y,m_i^x)$. Note that higher vortex numbers $|\nu_{ijk}|>1$ are energetically disfavored [cf. Eq.~(\ref{eq:vortex_energy})] and are not observed in our simulations. With this, the averaged number of vortex-antivortex pairs is computed as
\begin{equation}\label{eq:vortex_pair_number}
    N_v = \frac{1}{2}\left\langle\sum_{\langle ijk\rangle} |\nu_{ijk}| \right\rangle~,
\end{equation}
which takes into account that the number of vortices and antivortices has to be equal in order to fulfill the periodic boundary condition. The thermally averaged vortex-pair number obtained from this procedure is shown in Fig.~\ref{Fig4}(b), and the identification of vortices from a single $40\times40$ spin configuration is exemplarily illustrated for $T=149~$K [Fig.~\ref{Fig4}(c)] and $T=190~$K [Fig.~\ref{Fig4}(d)].

In the spin texture at $T=190~$K some unbound vortex-antivortex pairs can be identified. However, the dominant feature appears to be that the number of vortices increases strongly with temperature. A very important observation arises in comparison to the magnetization (black data points): vortex pair number and magnetization are inversely proportional. Moreover, the critical temperature from $M(T)/M_0$ seems to coincide with the steepest rise of the vortex number. This demonstrates that the thermodynamic anomaly in the easy-plane Heisenberg model -- the disagreement between transition temperatures and $T_{\text{BKT}}$ -- is instead accompanied by a rapid proliferation of vortex excitations. In this regard, we investigate the connection between vortex-pair nucleation and phase transitions in the next subsection.

\subsection*{E. Vortex analysis}
To provide a statistical measure of temperature-dependent vortex population, we chose an ansatz similar to Shannon entropy. In this regard, we define the entropy $S$ for ensembles of vortex-pairs as
\begin{equation}
    S(T)=-p(T)\ln\left[ p(T)\right]~, \quad p(T)=\frac{ N_v}{N}~,
\end{equation}
for simulation boxes with $N$ lattice sites. The probability $p(T)$ for an interstitial lattice site hosting a vortex core is thereby computed from the averaged number of vortices $N_v$. In particular, $S$ does not distinguish between bound and unbound vortex pairs and therefore characterizes vortex proliferation rather than vortex-antivortex unbinding itself.

The resulting entropy for $K=-50~$meV is shown in Fig.~\ref{Fig5}(a). Regarding the well-known relation to the heat capacity
\begin{equation}
    C_V = T\frac{\partial S}{\partial T}~,
\end{equation}
we expect the peak in $C_V$ to match the position of steepest ascent of $S$, its inflection point $T_I$. The physical origin of this relation is that at $T_I$ an overproportional amount of thermal energy can be deposited into the nucleation of vortex-antivortex pairs, whose costs are approximately described by Eq.~(\ref{eq:vortex_energy}), effectively increasing the heat capacity of the system. In order to determine this inflection point, we find that the entropy can be well described by the model function from Eq.~(\ref{eq:fit_magnetization}), mirrored at $T=T_c$. The inflection point is consequently extracted from the fit as $T_I=T_c+\delta T$ and indicated by a vertical line in Fig.~\ref{Fig5}(a). 
In particular, for $K=-50~\mathrm{meV}$, we obtain $T_I=184.6\pm1.5~\mathrm{K}$, which is close to the characteristic temperature of $T_c \approx 179~\mathrm{K}$ obtained independently from $M$ and $C_V$ [cf. Fig.~\ref{Fig2}(b) and (d)].

A quantitative comparison between characteristic temperatures obtained from various observables over $K\in[-50,50]~$meV can be found in Fig.~\ref{Fig5}(b). In the easy-axis regime, $K>0~$meV, the characteristic temperatures extracted from $C_V$ and $M$ are in close agreement with each other and evolve toward the triangular-lattice Ising limit $T_{\text{Ising}}=4J/(k_{\text{B}}\ln 3)= 211.3~\mathrm{K}$ \cite{okabe2025bkt, husimi1950statistics} with increasing $K$. However, a noticeable deviation from the Ising limit remains even at $K/J=10$, consistent with an estimation from the random phase approximation \cite{torelli2019calculating}, suggesting that the Ising limit is not reached for such a finite anisotropy. 

In the easy-plane regime, $K<0~$meV, the vortex-entropy inflection point $T_I$ follows the thermodynamic anomaly extracted from $C_V$ and $M$ in astonishing accuracy. This correspondence suggests that phase transitions in the easy-plane Heisenberg model are determined by the nucleation of vortex pairs, in interplay with finite-size effects, rather than by the unbinding process in BKT theory. The relation to these pseudo-topological transitions is further underlined by the disagreement between values for characteristic temperatures $T_c$ and $T_{\text{BKT}}=153.1~$K [cf. Fig.~\ref{Fig2}(b)]. 
Thus, our results raise the necessity of sophisticated thermodynamic simulations over analytic models for the prediction of phase transitions in realistic finite-size 2D magnetic systems.

\section{CONCLUSIONS} 

In summary, we present a systematic study of finite-temperature magnetism in the anisotropic 2D Heisenberg model across the easy-plane, isotropic, and easy-axis regimes. Finite-size effects vary strongly across the different spin-symmetry regimes, being strongest in the isotropic Heisenberg case and becoming substantially weaker toward the easy-plane and easy-axis limits. In the isotropic regime, the finite-size magnetic ordering temperature is determined by the relation between the spin-correlation length and the system size. Increasing easy-axis anisotropy progressively drives the system toward Ising-like behavior, although noticeable deviations remain even at $K/J=10$ compared to the Ising limit. The easy-plane regime exhibits a more unexpected behavior, with the magnetization and heat capacity showing a common thermodynamic anomaly at a temperature above $T_{\text{BKT}}$, obtained by extrapolating the finite-size spin-stiffness universal-jump crossings to the thermodynamic limit. Moreover, the sharp heat-capacity anomaly contrasts with the broad maximum typically associated with BKT behavior. Instead, the thermodynamic anomaly closely follows the rapid proliferation of vortices. We relate this observation to the ability of easy-plane magnets to deposit thermal energy into the nucleation of vortex-antivortex pairs, which effectively raises their heat capacity and explains the thermodynamic anomaly. Thus, our study reveals the relevance of the combined effect of finite-size correlations and vortex excitations rather than directly identifying pure BKT behavior in easy-plane Heisenberg magnets. Overall, our results identify magnetic anisotropy, spin symmetry, and system size, in interplay with their effect on vortex-antivortex pair proliferation, as key ingredients for the design of experimentally and technologically relevant temperature scales in 2D magnets.

\section*{APPENDIX A: SIMULATION DETAILS}

MC simulations were used to determine the temperature dependence of the
thermodynamic observables. We employed the Metropolis algorithm with adaptive single-spin sampling~\cite{alzate2019optimal}, as implemented in the \textsc{spinaker} code.
At each temperature, $10^4$ MC cycles were used for thermalization, followed by another $10^4$ cycles for evaluating thermodynamic averages and fluctuations of the internal energy $\langle E\rangle$, the magnetization $M$ [cf. Eq.~(\ref{eq:magnetization})] and the heat capacity $C_V$ [cf. Eq.~(\ref{eq:heat_capacity})]. From the stagnation of internal energy during the thermalization process, it was ensured that the system reached thermal equilibrium at each temperature. Other thermodynamic observables such as the spin stiffness [cf. Eq. (\ref{eq:spin_stiffness})], the number of vortex-antivortex pairs $N_v$ [cf. Eq.~(\ref{eq:vortex_pair_number})] and the correlation length (see Appendix B) have been averaged from 100 thermally averaged spin configurations at each temperature.

\section*{APPENDIX B: CORRELATION LENGTH}

To characterize the spatial correlations in the easy-plane regime, we compute the spin correlation function and subsequently the correlation length on the easy-plane components only. In this case, the correlator at each temperature reads
\begin{equation}
    %G_{c}(r)= \frac{1}{2}\sum_{\mu=1}^{2} \left\langle m_i^x m_{i+r\mathbf a_\mu}^x+ m_i^y m_{i+r\mathbf a_\mu}^y \right\rangle_i ,
    G(r,T) = \frac{1}{2} \left\langle\sum_{i,j} m_i^x m_j^x+ m_i^y m_j^y\right\rangle~,\quad |\mathbf{r}_i-\mathbf{r}_j|=r
\end{equation}
where 
$\langle\cdots\rangle$ represents an average over 100 thermalized spin configurations in simulation boxes with periodic boundary conditions.

The estimation of the correlation length $\xi$ is based on the prediction of BKT theory regarding the algebraic decay of correlations below and exponential decay above the critical temperature $T_{\text{BKT}}$ \cite{kosterlitz1973ordering,
Kosterlitz1974},
\begin{equation}
    G(r,T)\sim \begin{cases}
        r^{-\eta(T)}~, & T<T_{\text{BKT}}~,\\[4pt]
        \exp[-r/\xi(T)]~, & T\ge T_{\text{BKT}}~,
    \end{cases}
\end{equation}
where $\eta$ is some function of $T$. Thus, we estimate $\xi(T)$ by a least-square fit of the model function
\begin{equation}
    \Tilde{G}(r; A,\xi,c) = A\exp(-\frac{r}{\xi})+c~,
\end{equation}
to numerically obtained $G(r,T)$ for each temperature $T$ and $r$ in the range $2\le r/a\le18$.
The constant $c$ accounts phenomenologically for residual finite-size contributions to the correlation function. Subsequently, the temperature dependence of the correlation length $\xi$ is obtained by a least-square fit of Eq.~(\ref{eq:bkt_correlation_length}), as illustrated in Fig.~\ref{Fig3}(a). Note that the exponential behavior, by theory, is only valid in the regime $T>T_{\text{BKT}}$, but in our simulations, it is found to be further limited by finite-size effects (cf. discussion in Sec.~III.B). 

\section*{APPENDIX C: SPIN STIFFNESS}
For transparency, we provide the derivation and numerical implementation of the spin stiffness in Eq.~(\ref{eq:spin_stiffness}). The spin stiffness $\rho_s$
measures the quadratic response of the free energy $F$ to an infinitesimal
twist $\mathbf{q}\to0$ of the $xy$-components of spin configuration. 
Based on the Hamiltonian in Eq.~(\ref{eq:hamiltonian}), the action of a twist on the exchange interaction can be incorporated by the following modification
\begin{equation}\label{eq:twisted_hamiltonian}
\begin{aligned}
    \Tilde{H}(\mathbf q)= -J\sum_{\langle ij\rangle}\Big[ m_i^z m_j^z + 
&(m_i^x m_j^x+m_i^y m_j^y) \cos(\mathbf q\cdot \mathbf{r}_{ij})\\
+&(m_i^x m_j^y-m_i^y m_j^x) \sin(\mathbf q\cdot\mathbf{r}_{ij})
\Big]~.
\end{aligned}
\end{equation}
Note that this Hamiltonian recovers the original one in the limit $\mathbf{q}=0$. Further, considering the free energy $F(\mathbf{q})$ from the partition function $Z(\mathbf{q})$ over the manifold $\Omega=\bigotimes_{i=1}^N\mathbb{S}^2$ of spin configurations on a lattice with $N\in\mathbb{N}$ lattice sites
\begin{equation}
    F(\mathbf{q})=-\frac{1}{\beta}\ln Z(\mathbf{q})~,\quad Z(\mathbf{q})= \int_{\Omega} \mathrm{e}^{-\beta \Tilde{H}(\mathbf{q})}~\mathrm{d}\Omega~,
\end{equation}
its derivative with respect to $q_{\alpha}$, $\alpha\in\{x,y\}$, can be expressed as an ensemble average
\begin{equation}
    \begin{split}
        \frac{\partial F(\mathbf q)} {\partial q_\alpha} &= -\frac{1}{\beta Z(\mathbf{q})}\frac{\partial Z(\mathbf q)} {\partial q_\alpha}\\
        &= \frac{1}{Z(\mathbf{q})}\int_{\Omega} \frac{\partial \Tilde{H}(\mathbf q)} {\partial q_\alpha}\mathrm{e}^{-\beta \Tilde{H}(\mathbf{q})}~\mathrm{d}\Omega = \Big\langle\frac{\partial \Tilde{H}}{\partial q_{\alpha}}\Big\rangle ~.
    \end{split}
\end{equation}
Similarly one can write down the spin stiffness, which is the second derivative of $F(\mathbf{q})$ with respect to $q_{\alpha}$, as~\cite{obuchi2012spin}
\begin{equation}\label{eq:spin_stiffness_prototypical}
    \begin{split}
        \rho_s &= \frac{1}{NA} \left. \frac{\partial^2F(\mathbf q)} {\partial q_\alpha^2} \right|_{\mathbf q=0}\\
        &= \frac{1}{NA}\left.\left[\Big\langle\frac{\partial^2 \Tilde{H}}{\partial q_{\alpha}^2}\Big\rangle - \beta\Big[\Big\langle\Big(\frac{\partial \Tilde{H}}{\partial q_{\alpha}}\Big)^2\Big\rangle - \Big\langle\frac{\partial \Tilde{H}}{\partial q_{\alpha}}\Big\rangle^2\Big]\right]\right|_{\mathbf{q}=\boldsymbol{0}}~,
    \end{split}
\end{equation}
with $A=\sqrt{3}a^2/2$ as the unit cell area of the hexagonal lattice. Note that the result is independent of the choice of $\alpha$ in isotropic systems. From this form, it is easy to see that defining the observables $B_{\alpha}, I_{\alpha}\in\mathbb{R}$ as
\begin{equation}\label{eq:stiffness_terms}
    \begin{split}
        I_{\alpha} &:= \left.\frac{\partial \Tilde{H}}{\partial q_\alpha} \right|_{\mathbf{q}=\boldsymbol{0}} = J\sum_{\langle ij\rangle}r_{ij}^{\alpha} (m_i^xm_j^y - m_i^ym_j^x)~, \\
        B_{\alpha} &:= \left.\frac{\partial^2 \Tilde{H}}{\partial q_\alpha^2} \right|_{\mathbf{q}=\boldsymbol{0}} = J\sum_{\langle ij\rangle}(r_{ij}^{\alpha})^2 (m_i^xm_j^x + m_i^ym_j^y)~,
    \end{split}
\end{equation}
the expression for the spin stiffness in Eq.~(\ref{eq:spin_stiffness}) directly arises from the general form of the stiffness in Eq.~(\ref{eq:spin_stiffness_prototypical}). However, it is important to note that in this way the spin stiffness does not only contain the curvature of the energy with respect to a twist $\mathbf{q}$, which is described by $B_{\alpha}$, but also the reduction by thermal fluctuations of the spin current, described by $I_{\alpha}$.

	%%%%%%%%%%%%%%%%% Acknowledgement %%%%%%%%%%%%%%%%%%%%%%%
\section*{ACKNOWLEDGMENTS} 

This study has been supported France 2030 government investment plan managed by the French National Research Agency under grant reference PEPR SPIN – [SPINTHEORY] ANR-22-EXSP-0009. This study has been supported by the National Natural Science Foundation of China (Grant No. 12274228). This study has been (partially) supported through the grant NanoX no.~ANR-17-EURE-0009 in the framework of the ``Programme des Investissements d’Avenir". This work was performed using HPC resources from CALMIP (Grant No. 2024/2026-[P21023]). We thank T. Olsen for making us aware of the question regarding the existence of BKT transitions in the Heisenberg model and L. Kollwitz for technical support regarding the \textsc{spinaker} code.
		
	\bibliography{References}
	
\end{document}